\documentclass[onecollarge,natbib]{svjour2}
\bibpunct{[}{]}{;}{n}{}{,} % to get "[numbered]" references from natbib
\smartqed  % flush right qed marks, e.g. at end of proof
\usepackage{graphicx}
\usepackage{amsmath}
\newcommand{\half}{\tfrac{1}{2}}
\journalname{Few-Body Systems}
\begin{document}

\title{Multiparton interactions and multiparton distributions in
  QCD\thanks{Presented at LIGHTCONE 2011, 23 -- 27 May, 2011, Dallas.}
}
%\subtitle{Do you have a subtitle?\\ If so, write it here}

%\titlerunning{Short form of title}        % if too long for running head

\author{\begin{flushright}{\small DESY 11-195}\end{flushright}
  Markus Diehl
}

%\authorrunning{Short form of author list} % if too long for running head

\institute{Markus Diehl \at
              Deutsches Elektronen-Synchroton DESY, 22603 Hamburg, Germany \\
              \email{markus.diehl@desy.de}           %  \\
}

\date{Received: date / Accepted: date}
% The correct dates will be entered by the editor

\maketitle

\begin{abstract}
  After a brief recapitulation of the general interest of parton
  densities, we discuss multiple hard interactions and multiparton
  distributions.  We report on recent theoretical progress in their QCD
  description, on outstanding conceptual problems and on possibilities to
  use multiparton distributions as a laboratory to test and improve our
  understanding of hadron structure.
\end{abstract}

%%%%%%%%%%%%%%%%%%%%%%%%%%%%%%%%%%%%%%%%

\section{Parton distributions: a brief recapitulation}
\label{sec:motivation}

Among the most interesting and most difficult aspects of QCD is the
relationship between quarks and gluons, which are the basic degrees of
freedom of the theory, and hadrons, which are the physical states observed
at macroscopic distances.  Parton distributions are among the most
prominent quantities that describe this relationship.  Important issues in
this context are:
\begin{itemize}
\item Parton densities describe quarks, antiquarks and gluons as they
  manifest themselves in short-distance processes, and they have a direct
  connection to the fields that appear in the QCD Lagrangian.  How are
  these degrees of freedom related with the quarks that appear in
  non-perturbative approaches to hadron structure, such as constituent
  quark models, chiral quark models, the Dyson-Schwinger approach etc.?
\item The appearance of partons at small momentum fraction $x$ can to a
  large extent be described as the result of parton radiation in the
  perturbative regime, as it is for instance encoded in the DGLAP
  equations.  However, in particular the fits performed by the Dortmund
  group \cite{Gluck:1998xa,JimenezDelgado:2008hf} clearly show that
  perturbative physics alone is not sufficient for understanding sea
  quarks and gluons, given that they must already be present at low
  resolution scales, where they can only be of non-perturbative origin.
  Distributions such as $\bar{u}-\bar{d}$ or $s - \bar{s}$ are
  particularly clear indicators of this fact.
\item Which roles do confinement and chiral symmetry breaking play in
  determining the distribution of partons inside a hadron?
\item Gluons and the choice of gauge play an essential role when one
  describes the dynamics of partons inside hadrons.  Our understanding of
  this issue has seen important progress in the last decade but is still
  far from satisfactory, as we will see at the end of this section.
\end{itemize}
Parton distributions come in different varieties, which address
different aspects of hadron structure.
\begin{description}
\item[PDFs,] i.e.\ parton distribution functions $f(x)$, are the familiar
  functions describing the distribution of partons in longitudinal
  momentum fraction $x$.  There is a vast number of processes where they
  can be measured.  The PDFs for unpolarized partons in the proton are an
  indispensable ingredient for understanding high-energy lepton-hadron and
  hadron-hadron collisions, and with the exception of certain
  distributions they are among the most precisely measured
  non-perturbative quantities in QCD.
\item[TMDs,] i.e.\ transverse-momentum dependent distributions
  $f(x,\vec{k})$, describe the joint distribution of partons in their
  longitudinal momentum fraction and their transverse momentum.  They are
  relevant in processes with a measured transverse momentum much smaller
  than the hard scale, e.g.\ in Drell-Yan production when the transverse
  momentum of the lepton pair is much smaller than its invariant mass.
  (Here and in the following, vectors in the transverse plane are written
  in boldface.)

  TMDs quantify a variety of spin-orbit correlations at the parton level.
  A particularly intriguing aspect of their dynamics is the role of gluons
  that are described by Wilson lines (see below).  We note that the term
  TMD also includes transverse-momentum dependent fragmentation functions.
\item[GPDs,] i.e.\ generalized parton distributions $F(x,\xi,t)$, appear
  in the description of exclusive processes like deeply virtual Compton
  scattering or meson production.  The Mandelstam variable $t$ can be
  traded for the transverse momentum difference $\vec{\Delta}$ between the
  incident and the scattered proton, which is Fourier conjugate to the
  transverse position $\vec{b}$ of the parton inside the proton.  If the
  skewness variable $\xi$ (which describes the longitudinal momentum
  transfer to the target) is zero, the Fourier transformed distributions
  $f(x,\vec{b})$ describe the joint density of partons in their
  longitudinal momentum fraction and their transverse position
  \cite{Burkardt:2002hr}.  One often refers to $f(x,\vec{b})$ as impact
  parameter distributions.

  The Mellin moments $\int dx\, x^{n-1} F(x,\xi,t)$ of GPDs can be
  identified with the form factors of local operators.  This provides in
  particular a connection between GPDs and the electromagnetic Dirac and
  Pauli form factors of the nucleon, which have been measured with great
  precision over a wide range of $t$.  The moments of GPDs can also be
  evaluated in lattice QCD \cite{Hagler:2009ni}.
\end{description}
A common feature of these different types of distributions is that they
are defined in terms of matrix elements
\begin{equation}
  \label{gen-mat-el}
\langle p_2 |
  \bar{q}_\beta(z_2)\, W(z_2, z_1)\, q_\alpha(z_1) | p_1 \rangle
\Big|_{z_1^+ = z_2^+ = 0}
\end{equation}
of quark-antiquark operators between proton states.  For gluon
distributions, the fields $\bar{q}$ and $q$ are replaced by the gluon
field strength $F^{+i}$.  These matrix elements have a rather simple
interpretation in light-cone quantization, where the fields $\bar{q}(z_2)$
and $q(z_1)$ at zero light-cone time can be expanded in terms of creation
and annihilation operators for quarks or antiquarks (see e.g.\
\cite{Collins:2008ht}).  A subtle point of interpretation arises due to
the Wilson line $W(z_2, z_1)$, which will be discussed shortly.

Longitudinal and transverse coordinates play quite different roles in
the matrix element \eqref{gen-mat-el}, which means that one has lost manifest
three-dimensional rotation invariance even if both proton states $p_1$ and
$p_2$ are at rest.  This corresponds to the fact that in processes where
parton distributions can be observed, there is a physically preferred
direction.  The variables of $z_1^-$ and $z_2^-$ in \eqref{gen-mat-el} are
related to the light-cone plus-momenta of the quark or antiquark by a
Fourier transform, and a parton interpretation is most straightforward in
a reference frame where the protons move fast, i.e.\ where $p_1^+$ and
$p_2^+$ are large.  The transverse variables in \eqref{gen-mat-el} differ
between PDFs, TMDs and GPDs.  One can switch between transverse position
and transverse momentum of the proton states or the partons by a Fourier
transform.  There is a subtle point in this, which can easily be explained
by a simple calculation.  The Fourier transformed quark field
\begin{align}
q(z^-, \vec{k}) &= \int d^2\vec{z}\;
   e^{i \vec{k} \vec{z}}\, q(z^-, \vec{z})
\end{align}
at $z^+=0$ annihilates quarks with transverse momentum $\vec{k}$ and
creates antiquarks with transverse momentum $-\vec{k}$.  For a bilinear
operator as in \eqref{gen-mat-el} we then have
\begin{align}
  \label{FT-operator}
\bar{q}(\vec{k})\, q(\vec{k}')
 &= \int d^2\vec{z}\, d^2\vec{z}' \;
    e^{-i (\vec{k} \vec{z} - \vec{k}' \vec{z}')}\,
    \bar{q}(\vec{z})\, q(\vec{z}') \,,
\end{align}
where here and below we suppress plus- and minus-coordinates for
simplicity.  With
\begin{align}
\vec{k} \vec{z} - \vec{k}' \vec{z}'
 &= \frac{\vec{k} + \vec{k}'}{2}\, (\vec{z} - \vec{z}')
  + (\vec{k} - \vec{k}')\, \frac{\vec{z} + \vec{z}'}{2}
\end{align}
we see that in the region where $\bar{q}$ creates and $q$ annihilates a
quark, the \emph{average} transverse momentum of the created and
annihilated quark is Fourier conjugate to the \emph{difference} of their
transverse positions, whereas the difference of the transverse momenta is
conjugate to their average transverse position.  Taking the matrix element
of \eqref{FT-operator} between proton states that are localized at
transverse position $(\vec{z} + \vec{z}')/2 -\vec{b}$, and making an
appropriate Fourier transformed with respect to the minus-positions of the
fields, one obtains Wigner distributions $W(\vec{k}, \vec{b})$ that depend
on both the average transverse momentum of a quark and its average
transverse position relative to the proton, where the ``average'' refers
to the two fields in the operator~\eqref{FT-operator}.  The integral $\int
d^2\vec{b}\, W(\vec{k}, \vec{b})$ gives the density of quarks with
transverse momentum~$\vec{k}$, i.e.\ a TMD, whereas the integral $\int
d^2\vec{k}\, W(\vec{k}, \vec{b})$ gives the density of quarks with
transverse position~$\vec{b}$ in the proton, i.e.\ essentially a Fourier
transformed GPD.  This shows explicitly that TMDs and GPDs contain
complementary information about partons in the proton, and that both types
of quantities are descendants of higher-level functions, namely of Wigner
distributions.  For a general introduction to Wigner distributions we
refer to~\cite{Hillery:1984}.

\begin{figure}
\begin{center}
\includegraphics[width=\textwidth]{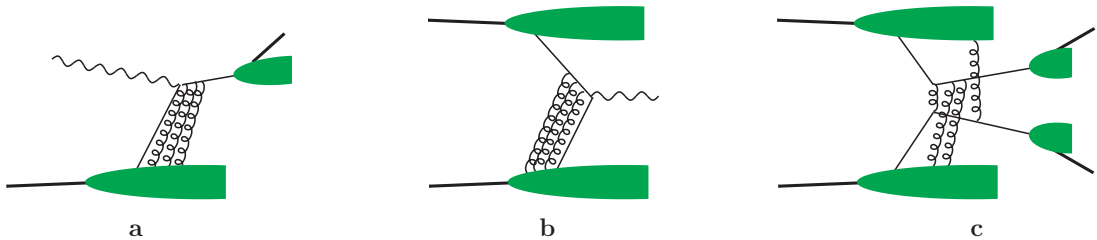}
\end{center}
\caption{\label{fig:tmd-fact} Graphs with gluon exchange between hadrons
  and the hard-scattering subprocess for semi-inclusive deep inelastic
  scattering (a), Drell-Yan production (b) and hadron pair production in
  $pp$ collisions (c).}
\end{figure}

Let us now return to the Wilson line in \eqref{gen-mat-el}, which we have
glossed over so far.  When establishing factorization for processes such
as Drell-Yan production or semi-inclusive deep inelastic scattering
(SIDIS), one finds that an arbitrary number of gluons in a right-moving
hadron can attach to the hard-scattering process without any twist
suppression, provided that the gluons have polarization in the
plus-direction.  This is illustrated in Fig.~\ref{fig:tmd-fact}a and b.
It is the effect of these gluons that is summed up in the Wilson line
$W(z_2, z_1)$, with a path between $z_1$ and $z_2$ that depends on the
process.

For $\vec{k}$ integrated distributions, the situation simplifies if one
works in the light-cone gauge $A^+ = 0$, where the Wilson line reduces to
unity so that one has a literal parton model interpretation of the
operator in \eqref{gen-mat-el} as creating and annihilating quarks or
antiquarks.  For $\vec{k}$ dependent distributions, however, there are
subtle spin effects such as the Sivers or Boer-Mulders asymmetries, which
are quite explicit in a covariant gauge, whereas in $A^+ = 0$ gauge
they are due to Wilson line pieces at infinity, corresponding to gluon
modes with $k^+ = 0$, see e.g.~\cite{Collins:2002kn,Belitsky:2002sm}.

Closer inspection shows that in TMDs one can actually not take the Wilson
line $W(z_2, z_1)$ along a path in the light-like direction, since this
leads to rapidity divergences \cite{Collins:2008ht}.  To regulate these
divergences, one can introduce a direction $v$ with nonzero plus- and
minus-components, so that the resulting Wilson lines reduce to unity in
the axial gauge $A v = 0$ rather than in light-cone gauge.  A formulation
of factorization with TMDs in the framework of light-cone quantization and
the $A^+=0$ gauge has so far not been achieved.  On the positive side, the
formulation with Wilson lines along a direction $v$ allows one to resum
Sudakov logarithms to all orders.  This has led to a successful
phenomenology of processes with small measured transverse momentum, see
\cite{Nadolsky:2001sf,Aybat:2011zv} and references therein.  On a more
fundamental level, the physical phenomena associated with TMDs force (and
guide) us to think more deeply about the role of gluons and of gauge
degrees of freedom when describing how parton emerge from hadrons and
interact with each other.

Finally, we must note that TMD factorization has so far only been
established for processes with a simple color structure in the hard
scattering, such as SIDIS and Drell-Yan production.  Serious obstacles for
such a formulation have been identified for processes like 
the production of back-to-back hadrons or jets in $pp$ collisions, where
graphs like the one in 
Fig.~\ref{fig:tmd-fact}c do not appear to admit a resummation into Wilson
line operators \cite{Rogers:2010dm}.  It remains an open question whether a
factorized description (and along with it, Sudakov resummation with the
same accuracy as for Drell-Yan production) can be found for this important
class of processes.

%%%%%%%%%%%%%%%%%%%%%%%%%%%%%%%%%%%%%%%%%%%%%%%%%

\section{Multiparton interactions: what are they and why are they
  interesting?}

\begin{figure}
\begin{center}
\includegraphics[width=0.85\textwidth]{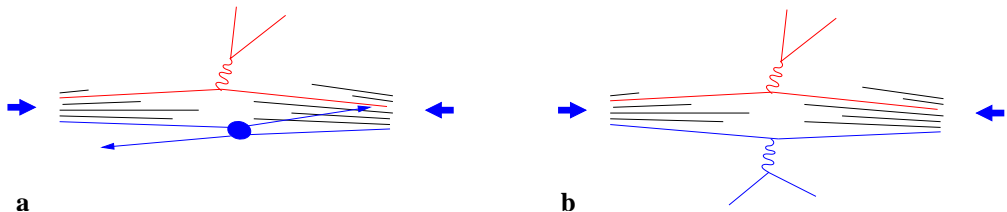}
\end{center}
\caption{\label{fig:double-dy} Double hard scattering processes producing
  a Drell-Yan lepton pair and a dijet (a) or two Drell-Yan pairs (b).}
\end{figure}

In a generic hadron-hadron collision, several partons in one hadron
scatter on corresponding partons in the other hadron.  At high energy,
several of these scatters can have a hard scale and produce particles with
large invariant mass or large transverse momenta.  Examples are shown in
Fig.~\ref{fig:double-dy}.  The effects of such multiparton interactions
(also referred to as multiple interactions) average out or are power
suppressed in sufficiently inclusive observables.  This is why they are
not included in the familiar factorization formulae for hadron-hadron
collisions, which involve single-parton densities and a single
hard-scattering subprocess.  However, for more exclusive observables and
more specific kinematics, contributions from multiple interactions can be
substantial, and 
it has been estimated that they will play an important  role in many
analyses at the LHC.  There is a long history of theoretical and
experimental studies of such interactions, and the field has seen a
strong boost in activity since the LHC has started operation, see e.g.\
the recent proceedings \cite{Bartalini:2010su}.

The description of multiparton interactions involves multiparton
distributions, which are interesting in their own right since they contain
information about correlations between different partons in the proton
wave function.  This information is inaccessible in the singe-parton
distributions we discussed in Sect.~\ref{sec:motivation}.

The existing phenomenology of multiple interactions (including
implementations in Monte Carlo event generators) is based on a simple and
physically intuitive picture, which involves however many simplifying
assumptions.  A systematic treatment in QCD, which would match the level
of sophistication achieved for single hard-scattering processes, remains
to be given.  Some steps into this direction have been taken in
\cite{Diehl:2011tt}, and in the present proceedings we will point out
results that have been obtained, open problems that have been identified,
and prospects of using multiple interactions to learn more about hadron
structure at the parton level and about theoretical approaches to 
understand this structure.

%%%%%%%%%%%%%%%%%%%%%%%%%%%%%%%%%%%%%%%%%%%%%%%%%

\section{Some basic results}
\label{sec:basic}

To begin with, let us specify the theoretical framework we will be using.
We consider double hard scattering as the simplest (and often most
prominent) case of multiparton interactions.  Requiring both parton-level
scattering processes to be hard allows us to use the concept of
factorization and the predictive power of perturbation theory.  Since the
phenomenological interest in multiple interactions comes from the
necessity to understand the hadronic final state in detail, we keep the
transverse momenta of the produced particles differential and are hence
led to using transverse-momentum dependent factorization and the associated
multiparton distributions.  Given the status of TMD factorization for
single hard scattering described above, we consider the double Drell-Yan
process of Fig.~\ref{fig:double-dy}b as a case where the prospects for
obtaining rigorous theory results are best, referring the practically
important cases where secondary interactions produce jets to future work.

\begin{figure}
\begin{center}
\includegraphics[height=0.28\textwidth]{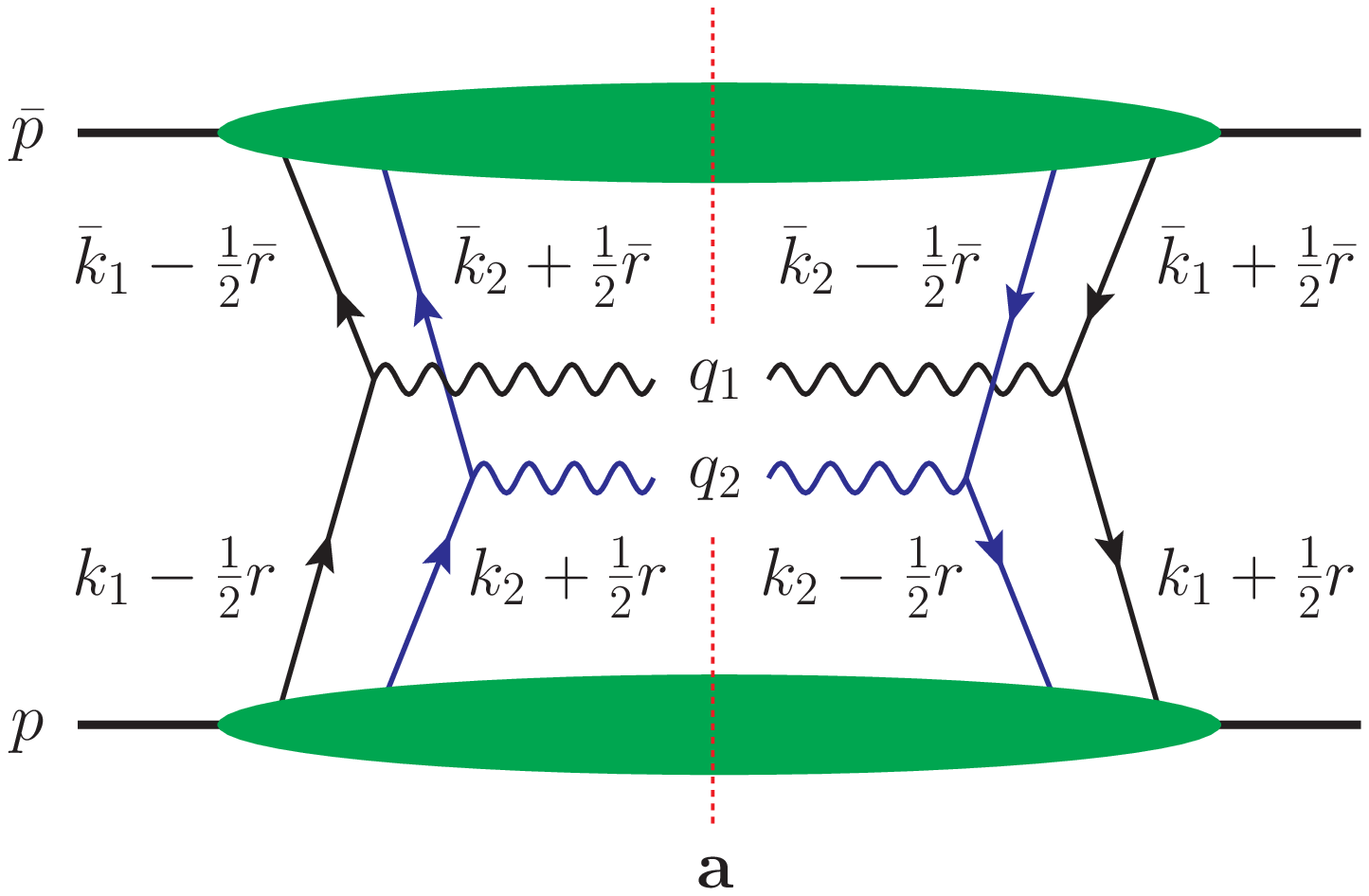}
\hspace{2em}
\includegraphics[height=0.28\textwidth]{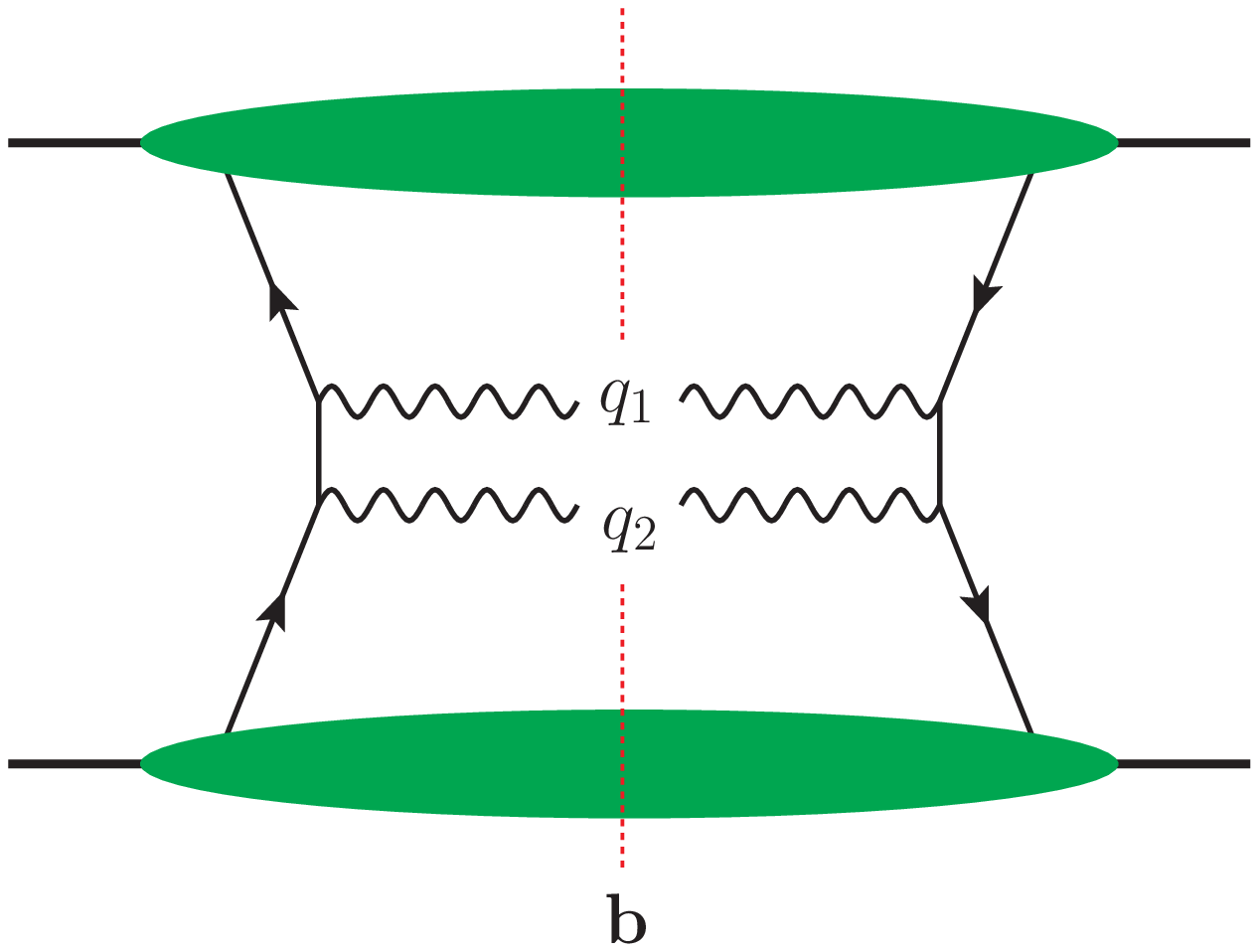}
\end{center}
\caption{\label{fig:double-single} Graphs for the production of two
  electroweak gauge bosons by double (a) and by single (b) hard
  scattering.  The dashed line indicates the final-state cut, and the
  decay of the bosons into leptons is not shown for simplicity.}
\end{figure}

The graph in Fig.~\ref{fig:double-single}a shows the production of a gauge
boson pair by two hard-scattering processes.  This graph can be evaluated
with the same techniques that are used for single hard scattering, where
small momentum components are neglected compared to large ones in each
parton-level subprocess.  In this approximation, the plus-momenta of
right-moving partons and the minus-momenta of left-moving ones are fixed
by the final-state kinematics, but their transverse momenta are not.  With
the momentum assignments in Fig.~\ref{fig:double-single}a, one finds
$x_i^{} = q_i^+ /p_{}^+$ and $\bar{x}_i^{} = q_i^- / \bar{p}_{}^-$ for the
longitudinal parton momentum fractions.  Momentum conservation gives
$\vec{r} + \bar{\vec{r}} = \vec{0}$ for the transverse components of the
momentum mismatch between partons on the left and the right of the
final-state cut.  The contribution of Fig.~\ref{fig:double-single}a to the
cross section is then proportional to $\int d^2\vec{r}\, F_{a_1, a_2}(x_i,
\vec{k}_i, \vec{r})\, F_{\bar{a}_1, \bar{a}_2}(\bar{x}_i, \bar{\vec{k}}_i,
-\vec{r})$, where the first factor is the distribution for two quarks and
the second factor the distribution for two antiquarks in the proton.
Fourier transforming the distributions w.r.t.\ their last argument, one
finds that the conjugate position variables satisfy $\bar{\vec{y}} =
\vec{y}$, and the full expression of the cross section reads
\begin{align}
  \label{X-section}
 \frac{d\sigma_{\,\text{Fig.~\protect\ref{fig:double-single}a}}}{%
  \prod_{i=1}^2 dx_i\, d\bar{x}_i\, d^2\vec{q}{}_i}
&= \frac{1}{S}
  \sum_{\substack{a_1, a_2 = q, \Delta q, \delta q \\[0.1ex]
          \bar{a}_1, \bar{a}_2 = \bar{q}, \Delta\bar{q}, \delta\bar{q}}}
  \biggl[\, \prod_{i=1}^{2} \,
  \int d^2\vec{k}_i\, d^2\bar{\vec{k}}_i\;
    \delta^{(2)}(\vec{q}{}_i - \vec{k}_i - \bar{\vec{k}}_i) \biggr]
\nonumber \\
& \quad \times  
  \hat{\sigma}_{1, a_1 \bar{a}_1}(q_1^2) \;
  \hat{\sigma}_{2, a_2 \bar{a}_2}(q_2^2)
\int d^2\vec{y}\,
  F_{a_1, a_2}(x_i, \vec{k}_i, \vec{y})\;
  F_{\bar{a}_1, \bar{a}_2}(\bar{x}_i, \bar{\vec{k}}_i, \vec{y}) \,,
  \phantom{\int}
\end{align}
where $\hat{\sigma}_{i, a_i \bar{a}_i}$ denotes the hard-scattering cross
section for single-boson production.  The statistical factor $S$ is $2$ if
the produced bosons are identical and $1$ if they are not.

From our discussion in Sect.~\ref{sec:motivation} it follows that the
Fourier conjugate variable $\vec{y}$ of $\vec{r}$ represents the distance
between the two scattering partons, averaged between the scattering
amplitude and its complex conjugate.  On the other hand, $\vec{k}_1$ and
$\vec{k}_2$ are the transverse momenta of the partons, averaged in the
same sense.  $F_{a_1, a_2}(x_i, \vec{k}_i, \vec{y})$ thus has the
structure of a Wigner distribution in the transverse degrees of freedom.
It is gratifying that this allows for a rather intuitive interpretation of
the cross section formula \eqref{X-section}: we have two quark-antiquark
annihilation processes separated by a transverse distance $\vec{y}$, where
in each annihilation the incident transverse parton momenta $\vec{k}_i$
and $\bar{\vec{k}}_i$ add up to the measured transverse momentum
$\vec{q}_i$ of the produced gauge boson.  Note, however, that the result
\eqref{X-section} is fully quantum mechanical and has been obtained from
Feynman graphs using standard approximations, without the need to appeal
to semi-classical arguments.  The operator definition of a two-quark
distribution reads
\begin{align}
  \label{dist-def}
F_{a_1,a_2}(x_i, \vec{k}_i, \vec{y})
&= \biggl[\, \prod_{i=1}^2
       \int \frac{dz_i^- d^2\vec{z}_i^{}}{(2\pi)^3}\,
       e^{i (x_i^{} z_i^- p^+ - \vec{z}_i^{} \vec{k}_i^{})}
    \biggr] \; 2 p^+\!\!\! \int dy^-
    \big\langle p \,\big|\,
    \bar{q}(- \half z_2)\, W(- \half z_2, \half z_2)\,
       \Gamma_{a_2} \, q(\half z_2) \;
\nonumber \\
 &\quad \times
    \bar{q}(y - \half z_1)\, W(y - \half z_1, y + \half z_1)\,
       \Gamma_{a_1} \, q(y + \half z_1)\,
    \big| \, p \big\rangle \Big|_{z_1^+ = z_2^+ = y^+ = 0}
\end{align}
and is a natural extension of the definition of TMDs for a single parton
described in Section~\ref{sec:motivation}.

The transverse-momentum integrated distribution $F_{a_1, a_2}(x_i,
\vec{y}) = \int d^2\vec{k}_1\, d^2\vec{k}_2\; F_{a_1, a_2}(x_i, \vec{k}_i,
\vec{y})$ gives the joint density of two quarks with momentum fractions
$x_1$ and $x_2$ and relative transverse distance~$\vec{y}$.  It enters in
the transverse-momentum integrated cross section as
\begin{align}
  \label{X-section-int}
& \frac{d\sigma_{\,\text{Fig.~\protect\ref{fig:double-single}a}}}{%
  \prod_{i=1}^2 dx_i\, d\bar{x}_i}
= \frac{1}{S}
  \sum_{\substack{a_1, a_2 = q, \Delta q, \delta q \\[0.1ex]
          \bar{a}_1, \bar{a}_2 = \bar{q}, \Delta\bar{q}, \delta\bar{q}}}
  \hat{\sigma}_{1, a_1 \bar{a}_1}(q_1^2) \;
  \hat{\sigma}_{2, a_2 \bar{a}_2}(q_2^2)
\int d^2\vec{y}\,
  F_{a_1, a_2}(x_i, \vec{y})\;
  F_{\bar{a}_1, \bar{a}_2}(\bar{x}_i, \vec{y}) \,.
\end{align}
The integration over $\vec{k}_1$ and $\vec{k}_2$ puts the relative
transverse coordinates $\vec{z}_1$ and $\vec{z}_2$ to zero in
\eqref{dist-def}, so that in each quark-antiquark operator the two fields
are separated by a light-like distance, as they are for the usual PDFs.
However, the operators describing the first and the second quark are still
separated by a transverse distance $\vec{y}$.  The resulting distribution
does therefore not involve a twist-four operator but rather the product of
two operators with twist two.

For each of the two quark-antiquark operators there are three Dirac
matrices $\Gamma_a$ that give a leading contribution to the cross section,
namely $\Gamma_{q} = \half \gamma^+$, $\Gamma_{\Delta q} = \half \gamma^+
\gamma_5$ and $\Gamma_{\delta q}^j = \half i \sigma^{j +} \gamma_5$ with
$j=1,2$.  These matrices respectively project on unpolarized,
longitudinally polarized and transversely polarized quarks and are
well-known from the definition of single-parton densities.  Note that for
multiparton distributions one can have spin effects even in an unpolarized
proton because the polarizations of different partons can be correlated
among themselves.  If such correlations are large, they leave an imprint
in the cross section formulae \eqref{X-section} and \eqref{X-section-int},
where they can change both the size of the cross section and the
distribution of particles in the final state.  In particular, one finds
that a correlation between the transverse polarization of two quarks or
antiquarks leads to an angular correlation between the leptonic decay
planes of the two produced gauge bosons.  Whether and where spin
correlations between different partons in the proton are important is
therefore an important question for phenomenology at the LHC.  It is also
of great interest from the point of view of hadron structure, and one may
hope for a fruitful interplay between these two fields in the future.

Not only the spin but also the color of two partons can be correlated.  In
Eqs.~\eqref{X-section} to \eqref{X-section-int} we have glossed over this
degree of freedom, and it turns out that two-quark distributions can have
two color structures, both of which contribute to the cross section.  The
two quark lines with momentum fraction $x_1$ in
Fig.~\ref{fig:double-single}a can couple either to a color singlet or to a
color octet.  A corresponding color coupling of the two quark lines with
momentum fraction $x_2$ then follows from color conservation.  It is the
color singlet combination that has the interpretation of a Wigner
distribution discussed above, whereas the color octet combination
describes an interference between different color states of a parton in
the scattering amplitude and its complex conjugate.  The operator defining
the color octet combination reads $(\bar{q}_2\, \Gamma_{a_2} t^a q_2)\,
(\bar{q}_1 \Gamma_{a_1} t^a q_1)$, where $t^a$ denotes $\half$ times the
Gell-Mann matrices.  The subscripts $1$ and $2$ of the quark fields
indicate that after Fourier transformation the fields are associated with
momentum fraction $x_1$ and $x_2$, respectively.
Although the possibility of spin and color correlations in multiparton
distributions has been pointed out long ago \cite{Mekhfi:1985dv}, the lack
of a method to estimate their size has prevented them from being included
in phenomenological analyses up to now.

In Sect.~\ref{sec:motivation} we mentioned the resummation of Sudakov
logarithms in single hard-scattering processes with small transverse
momenta.  Such logarithms also appear in the double scattering process
depicted in Fig.~\ref{fig:double-single}a.  They can be computed by
generalizing the method of \cite{Collins:1981uk,Collins:1984kg} from
single to double hard scattering.  An important result is that the leading
double logarithms in $\vec{q}_i^2 /Q^2$ are given by the product of the
corresponding factors for each separate hard-scattering process.  Beyond
this accuracy, the structure becomes more involved since soft gluons can
be exchanged between all parton lines in Fig.~\ref{fig:double-single}a.
Explicit calculation shows that these effects also lead to a mixing
between the color singlet and color octet distributions mentioned in the
previous paragraph.

%%%%%%%%%%%%%%%%%%%%%%%%%%%%%%%

\section{Power counting: when are multiple interactions suppressed and
  when are they not?}

When deriving the cross section formula \eqref{X-section} we have taken
the leading approximation in the small parameter $\Lambda /Q$, where the
large scale is provided by $Q^2 \sim q_1^2 \sim q_2^2$, whereas the small
scale $\Lambda$ is given by the transverse boson momenta $|\vec{q}_1|$,
$|\vec{q}_2|$ or by the scale of nonperturbative interactions, whichever
is larger.  One readily finds that, up to logarithmic terms, the overall
scaling behavior of the double-scattering cross section \eqref{X-section}
is
\begin{align}
  \label{power-double}
\frac{d\sigma_{\,\text{Fig.~\protect\ref{fig:double-single}a}}}{%
  \prod_{i=1}^2 dx_i\, d\bar{x}_i\; d^2\vec{q}_i}
& \sim\, \frac{1}{Q^4 \Lambda^2} \,.
\end{align}
One finds the same behavior for the case where the two bosons are produced
in a single hard scattering, as illustrated in
Fig.~\ref{fig:double-single}b.  In the fully differential cross section,
multiple hard interactions are thus \emph{not} power suppressed with
single hard scattering and can hence be of significant size.

The situation changes if one integrates over the transverse momenta
$\vec{q}_1$ and $\vec{q}_2$.  In the double scattering mechanism both
$|\vec{q}_1|$ and $|\vec{q}_2|$ result from transverse parton momenta and
are hence limited to size $\Lambda$.  By contrast, with a single hard
scattering the sum $|\vec{q}_1 + \vec{q}_2|$ is of order $\Lambda$ but the
individual transverse momenta can be as large as kinematically allowed,
i.e.\ they can be of order $Q$.  Single hard scattering can thus populate
a larger phase space, and one has
\begin{align}
\frac{d\sigma_{\,\text{Fig.~\protect\ref{fig:double-single}a}}}{%
  \prod_{i=1}^2 dx_i\, d\bar{x}_i}
& \sim\, \frac{\Lambda^2}{Q^4} \, , & 
\frac{d\sigma_{\,\text{Fig.~\protect\ref{fig:double-single}b}}}{%
  \prod_{i=1}^2 dx_i\, d\bar{x}_i}
& \sim\, \frac{1}{Q^2} \,.
\end{align}
The double scattering contribution is now power suppressed in the overall
cross section.  This is indeed necessary for the consistency of the usual
factorization formulae, which only describe the single scattering
contribution.

Multiple interactions are thus important in observables that are
sufficiently exclusive and in kinematics where this mechanism can
contribute.  The relevant kinematic regions are those where one has
subsets of final-state particles for which the vector sum of transverse
momenta is small compared with the hard scale in the process.  In our case
this is the double Drell-Yan process with small transverse momenta of the
produced vector bosons.  Another important example emphasized in
\cite{Blok:2011bu} is the production of two dijet pairs, each of which is
approximately back-to-back.

%%%%%%%%%%%%%%%%%%%%%%%%%%%%%%%

\section{High transverse momentum}

The predictive power of the theory increases if one considers kinematics
where the transverse momenta $\vec{q}_i$ are large compared with the scale
of nonperturbative interactions (while still being small compared with the
very large scale $Q$).  At least some of the transverse parton momenta are
then large as well, and one can compute the transverse-momentum dependent
multiparton distributions in terms of a perturbative subprocess at scale
$|\vec{q}_i|$ and distributions that depend on fewer variables.

\begin{figure}[b]
\begin{center}
\includegraphics[width=0.88\textwidth]{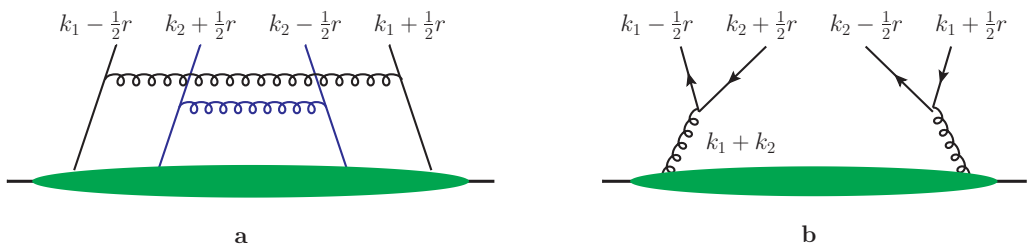}
\end{center}
\caption{\label{fig:high-qt} Graphs where partons with high transverse
  momentum are generated by parton radiation (a) or by the splitting of a
  single parton into two (b).}
\end{figure}

Two types of graphs are of particular importance in this context.  In the
double ladder graph of Fig.~\ref{fig:high-qt}a, each of the two partons
that will enter the collision process acquires a large transverse momentum
by radiating a parton into the final state.  This is a straightforward
generalization of the corresponding mechanism for single-parton densities
at large $\vec{k}$.  The two ladders in the graph are independent, so that
this mechanism can contribute for interparton distances $\vec{y}$ of
hadronic size.  An important finding is that the color factors for ladder
graphs favor the case where the parton pairs with equal momentum fraction
$x_1$ or $x_2$ are coupled to color singlets.  It remains to be studied
quantitatively how strong this preference for the color singlet channel
is.

\begin{figure}
\begin{center}
\includegraphics[width=0.45\textwidth]{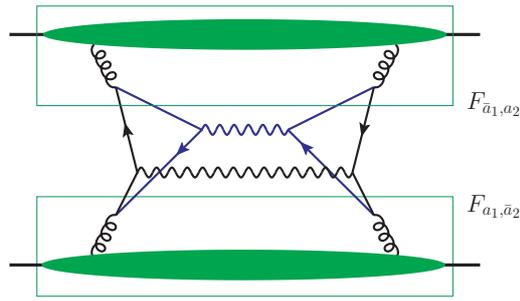}
\end{center}
\caption{\label{fig:high-X-sect} Graph for the production of two gauge
  bosons, where in both colliding protons an initial gluon splits into a
  quark-antiquark pair.  The boxes indicate the splitting graph of
  Fig.~\protect\ref{fig:high-qt}b for a quark-antiquark distribution.}
\end{figure}

In the graph of Fig.~\ref{fig:high-qt}a a single parton splits into two
partons with high transverse momenta, both of which will then take part in
a hard collision process.  This mechanism is in particular relevant for
large $\vec{r}$, which translates to a small distance $\vec{y}$.  One
finds that the transverse-momentum integrated distribution $F(x_i,
\vec{y})$ behaves like $1/\vec{y}^2$ at small $\vec{y}$ due to these
splitting graphs.  This poses a problem of consistency, since the integral
over $\vec{y}$ in the cross section \eqref{X-section-int} is then linearly
divergent in $\vec{y}^2$ at short distances.  Likewise, one finds that the
$\vec{y}$ integral in \eqref{X-section} diverges logarithmically at small
$\vec{y}$.  The underlying problem signaled by these divergences is that
the graph shown in Fig.~\ref{fig:high-X-sect} can be interpreted in two
ways.  If one identifies the two boxes in the figure as representing the
splitting graphs for a quark-antiquark distribution, one has a graph for
double hard scattering as just discussed.  However, without the boxes one
has a graph for single hard scattering, namely for the fusion of two
gluons into two electroweak gauge bosons via a quark loop.  For a detailed
investigation of this loop graph we refer to~\cite{Gaunt:2011xd}.  A
proper theoretical formulation will need to make sure that there is no
double counting problem for this graph and remove the divergences in the
$\vec{y}$ integration of the double scattering cross sections
\eqref{X-section} and \eqref{X-section-int}.  To achieve this in a
consistent and practicable way is an outstanding problem.

%%%%%%%%%%%%%%%%%%%%%%%%%%%%%%%

\section{Approximation by single-parton distributions}
\label{sec:gpd-connect}

To develop a phenomenology of multiple interactions, one needs a simple
ansatz for multiparton distributions as a starting point.  For those
distributions that admit an interpretation as Wigner distributions it is
natural to approximate them by the product of single-parton densities,
neglecting any possible correlations between the different partons.

\begin{figure}[b]
\begin{center}
\includegraphics[width=0.95\textwidth]{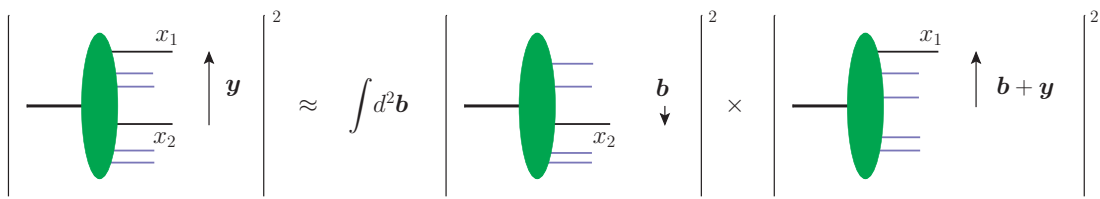}
\end{center}
\caption{\label{fig:approx} Visualization of the approximation
  \protect\eqref{fact-approx} of a two-parton distribution in terms of
  impact parameter densities for single partons.  In the figure we use
  that both types of distributions can be represented as squares of
  light-cone wave functions of the proton.}
\end{figure}

To formalize such an ansatz we insert a complete set of intermediate
states $\sum_X |X\rangle \langle X|$ between the two operators
$(\bar{q}_2\, \Gamma_{a_2} q_2)$ and $(\bar{q}_1\, \Gamma_{a_1} q_1)$ in
the definition \eqref{dist-def} of the two-parton distribution.  If we
\emph{assume} that this sum is dominated by single-proton states, then we
obtain an approximation in terms of single-parton distributions.  For
transverse-momentum integrated distributions, this approximation reads
\begin{align}
  \label{fact-approx}
F_{a_1, a_2}(x_i, \vec{y}) &\approx 
  \int d^2\vec{b}\; f_{a_2}(x_2, \vec{b})\,
                    f_{a_1}(x_1, \vec{b} + \vec{y}) \,,
\end{align}
where $f_a(x, \vec{b})$ is the impact parameter distribution of parton $a$
discussed in Sect.~\ref{sec:motivation}.  This approximation, visualized
in Fig.~\ref{fig:approx}, is the starting point of most estimates for
multiparton scattering in the literature, although there is a number of
attempts to go beyond it
\cite{Calucci:1999yz,Domdey:2009bg,Rogers:2009ke}.

The corresponding approximation for transverse-momentum dependent
distributions has a simpler structure if we use the variable $\vec{r}$
rather than $\vec{y}$.  We then have
\begin{align}
  \label{fact-approx-kt}
F_{a_1, a_2}(x_i, \vec{k}_i, \vec{r}) &\approx
  f_{a_2}(x_2, \vec{k}_2 - \half x_2 \vec{r}; -\vec{r})\,
  f_{a_1}(x_1, \vec{k}_1 - \half x_1 \vec{r}; \vec{r}) \,.
\end{align}
Here $f_a(x, \vec{k}, \vec{r})$ is a transverse-momentum dependent
generalized parton distribution, defined by the Fourier transform of a
matrix element as in \eqref{gen-mat-el}, where the annihilated quark
carries momentum fraction~$x$ and transverse momentum $\vec{k} - \half
\vec{r}$ and the created quark carries momentum fraction $x$ and
transverse momentum $\smash{\vec{k} + \half \vec{r}}$.

This strategy of approximation can also be adapted to the color octet
distributions described in Sect.~\ref{sec:basic}, which cannot be
interpreted as Wigner distributions.  To this end, we perform a Fierz
transformation in color and in spin space,
\begin{align}
(\bar{q}_2\, \Gamma_{a_2} t^a q_2)\, (\bar{q}_1 \Gamma_{a_1} t^a q_1)
 &= - \tfrac{1}{6}\, (\bar{q}_2\, \Gamma_{a_2} q_2)\,
                     (\bar{q}_1 \Gamma_{a_1} q_1)
    + \half \sum_{b_1, b_2} c_{a_1 a_2, b_1 b_2}\,
                     (\bar{q}_1 \Gamma_{b_2}\, q_2)\,
                     (\bar{q}_2\, \Gamma_{b_1}\, q_1) \,,
\end{align}
where the coefficients $c_{a_1 a_2, b_1 b_2}$ are straightforward to
calculate.  The first term on the r.h.s.\ is a color singlet distribution
and can be approximated as described above.  In the second term, the quark
and antiquark fields coupled to color singlets are now associated with
different longitudinal momentum fractions.  Repeating the steps that lead
to \eqref{fact-approx-kt} one finds that this term is represented by
transverse-momentum dependent GPDs in which not only the transverse but
also the longitudinal parton momenta differ between the created and the
annihilated quark.  This just means that the skewness variable $\xi$
mentioned in Sect.~\ref{sec:motivation} is nonzero in this case.

It should be emphasized that the above approximations are ad hoc in the
sense that we cannot quantify to which extent the sum over all
intermediate states $X$ in our derivation is actually dominated by
single-proton states.  However these approximations may serve as a
starting point, and they can give valuable information about aspects that
are poorly known otherwise, such as the distribution of partons in
transverse space and the possible size of color interference effects.

%%%%%%%%%%%%%%%%%%%%%%%%%%%%%%%

\section{Possible strategies for modeling multiparton distributions}

Our current knowledge of multiparton distributions is very limited, and it
should be interesting to investigate them in approaches that have
successfully been applied to single-parton densities.  The comparison of
predictions obtained in such approaches with upcoming data on multiple
interactions at LHC could lead to significant progress, with benefits for
both sides.

One possible avenue is to take Mellin moments of two-parton distributions
and to study the resulting matrix elements in lattice QCD as described in
\cite{Diehl:2011tt}.  This will naturally give limited information about
the dependence of the distributions on their momentum fractions $x_i$, but
it has the potential to quantify parton correlations in a genuinely
nonperturbative framework.

Constituent quark models, chiral quark models, the Dyson-Schwinger
approach and similar methods should be suitable to evaluate not only the
distributions of single partons but also two-parton distributions.  As
illustrated by simple examples in \cite{Diehl:2011tt}, one may expect
two-parton correlations of significant size in these approaches.  They can
also be used to investigate the degree of validity of the approximations
\eqref{fact-approx} and \eqref{fact-approx-kt}.

The two avenues just discussed will mostly give information about partons
with momentum fractions~$x_i$ above, say, $0.1$.  In multiple interactions
at LHC, typical values of $x_i$ are smaller by at least an order of
magnitude.  One way to investigate this region would be to take the
results from quark models as an input for scale evolution, which will
generate partons with lower $x_i$ by perturbative splitting as the scale
is increased.  Obvious questions to study in this context are to which
extent scale evolution may wash out two-parton correlations present at low
scales, and whether evolution to higher scales might improve the quality
of the approximation in \eqref{fact-approx}.  These would be important
lessons, even if one has to bear in mind that perturbative evolution alone
cannot account for sea quarks and gluons at small $x$, as already
mentioned in the beginning of Sect.~\ref{sec:motivation}.  An outstanding
challenge for the study of hadron structure is to find and develop
nonperturbative approaches suitable to describe these degrees of freedom
in a quantitative way.

%%%%%%%%%%%%%%%%%%%%%%%%%%%%%%%%%%%%%%%%%%%%%%%%%

\begin{acknowledgements}
  The results presented here have been obtained in collaboration with
  D.~Ostermeier and A.~Sch\"afer.  It is a pleasure to thank S.~Dalley
  and his colleagues for hosting a wonderful Light Cone meeting.
\end{acknowledgements}

%%%%%%%%%%%%%%%%%%%%%%%%%%%%%%%%%%%%%%%%%%%%%%%%%

% Non-BibTeX users please use

\end{document}